# Advancing ECG Diagnosis Using Reinforcement Learning on Global Waveform Variations Related to P Wave and PR Interval

Rumsha Fatima[1], Shahzad Younis[2], Faraz Shaikh[1], Hamna Imran[1], Haseeb Sultan[1], Shahzad Rasool[1] and Mehak Rafiq[1*]

[1] School of Interdisciplinary Sciences and Engineering, National University of Sciences and Technology, Islamabad, Pakistan
[2] School of Electrical Engineering and Computer Science, National University of Sciences and Technology, Islamabad, Pakistan

*Corresponding author: Mehak Rafiq (e-mail: mehak@sines.nust.edu.pk).

***Abstract*** — The reliable diagnosis of cardiac conditions through electrocardiogram (ECG) analysis critically depends on accurately detecting P waves and measuring the PR interval. However, achieving consistent and generalizable diagnoses across diverse populations presents challenges due to the inherent global variations observed in ECG signals. This paper is focused on applying the Q learning reinforcement algorithm to the various ECG datasets available in the PhysioNet/Computing in Cardiology Challenge (CinC). Five ECG beats, including Normal Sinus Rhythm, Atrial Flutter, Atrial Fibrillation, 1st Degree Atrioventricular Block, and Left Atrial Enlargement, are included to study variations of P waves and PR Interval on Lead II and Lead V1. Q-Agent classified 71,672 beat samples in 8,867 patients with an average accuracy of 90.4% and only 9.6% average hamming loss over misclassification. The average classification time at the 100th episode containing around 40,000 samples is 0.04 seconds. An average training reward of 344.05 is achieved at an alpha, gamma, and SoftMax temperature rate of 0.001, 0.9, and 0.1, respectively.
***Index Terms*** — P waves, PR Interval, Deep Learning, Reinforcement Learning, PhysioNet CinC, ECG Classification

## I. INTRODUCTION

THE Electrocardiogram (ECG) analysis has a long history that dates back to the first ECG recordings in the early 20th century. ECG has established itself as a crucial diagnostic tool for determining cardiac health throughout time. Despite its lengthy history, ECG delineation issues still exist, necessitating ongoing research and development to improve its diagnostic capabilities. This idea is backed up by studies[1], which emphasize that correct ECG analysis is still challenging due to interference from noise, artifacts, and individual variances in signal shape.

Researchers have been actively investigating novel ways to enhance ECG signal analysis and utilizing developments in filtering techniques, pattern recognition algorithms, and classification methods [2] to enhance ECG signal analysis. Recent improvements in processing speed and memory, alongside deep learning models, have enabled the processing of large ECG datasets, facilitated automatic feature extraction, and enhanced signal delineation. Convolutional neural networks (CNNs) have shown promise in identifying cardiac issues in ECG readings, outperforming traditional methods [3].

Accurately detecting cardiac problems based on the P waves and PR interval is one of the unexplored areas of ECG analysis. The PR interval indicates the conduction time from the SA node to the ventricles through the AV node, while the P wave reflects the depolarization of the atria [4]. Both characteristics significantly predict AV nodal conduction and atrial rhythm and function.

However, several variables that might impact P waves and PR intervals frequently make detecting and interpreting them difficult. P waves are often low-amplitude characteristics. Therefore, background noise in the signal can easily hide them [5]. This problem is especially prevalent in interpreting human illnesses Atrial Fibrillation (AF) [6], where variances in the diagnosis can result from variations in medical knowledge.

Additionally, human error and incorrect interpretations may occur when assessing additional cardiac conditions linked to P waves and PR intervals, such as atrial enlargement, junctional arrhythmias, and pre-excitation syndromes [7]. These conditions can change the P wave's shape, duration, or polarity and shorten or lengthen the PR interval.

The extensive nature of this challenge has prompted a widespread focus on analyzing beat-to-beat intervals using the QRS complex as a reliable fiducial point. To replicate the comprehensive clinical diagnostic procedure accurately, it is necessary to consider patient-specific history over a range of minutes to hours for a complete analysis and classification of global context disease patterns in the sequential data [8]. Reinforcement learning (RL) is a type of machine learning that helps improve decision-making by using experience and feedback from the environment [9]. Unlike other methods that rely on single-instance, exhaustive, and externally provided reward signals, RL deals with problems where decisions are made step-by-step and feedback is received over time. This makes RL worthwhile in healthcare areas with complex processes where diagnostic decisions and treatment protocols often encompass prolonged and sequential procedures [10].

By learning from experience and adjusting strategies, RL can develop robust solutions that make better decisions and improve ECG diagnosis outcomes. Electrocardiogram (ECG) diagnosis is one of the healthcare domains where RL can be utilized. An ECG monitors the heart's electrical activity and can identify several cardiac problems. However, ECG diagnosis is difficult because of the complexity and variability of the ECG signals, as well as the absence of consensus among experts and standardized criteria. Automated and intelligent approaches are required to help clinicians understand ECG data and provide precise and timely diagnoses [11].

The paper is organized as follows: Section 2 gives literature on traditional supervised learning and a new paradigm of reinforcement learning applications in ECG analysis. Section 4 provides a methodology and our proposed solution for generalizability across multi-institutional ECG datasets. Section 5 shares the results and insights of the proposed solution. Essential conclusions are summarized in Section 6.

## II. RELATED WORK

Using an electrocardiogram (ECG) to measure various cardiac characteristics is widespread usage. It is usually used in conjunction with a procedure that makes it simpler to record the electrical activity of the cardiac muscle over a predetermined time frame [12]. Throughout this process, several probes are placed in various spots to identify regions of a naked chest. These probes generate electricity by sensing each heartbeat's electrical activity on the chest's surface.

An ECG is commonly used in medicine to monitor the minute electrical changes in a patient's skin brought on by their heartbeat. This simple, non-invasive test successfully diagnoses a variety of heart conditions. The medical industry develops specific tools to help in diagnostics [13]. A high-resolution oscilloscope must capture the waveform for this gadget to display it. This strategy aims to create an effective P-QRS-T wave recognition-based ECG waveform categorization. This strategy will be focused on identifying the ECG waveform associated with cardiac functionality [14].

ECG analysis systems have historically applied well-known signal processing methods such as discrete wavelet transformations, dynamic mode decomposition, and principal component analysis to unstructured biological data. Recent research shows that categorizing patterns using machine learning increases the precision of arrhythmia detection and result interpretation [15].

Traditionally, clinical diagnosis in machine learning has been approached as a supervised classification problem [16], relying on many annotated samples to make predictions. Machine-learning models are trained using estimated peaks, durations between peaks, and other ECG signal features to automate the classification of heart disease.

Most studies in this field primarily utilize the MIT-BIH database, which is limited to data from only 48 patients [17]. For instance, the P wave detection algorithm in ECG signals, including pathological cases, was validated using three annotated databases: MITDB, QT database [18], and BUT PDB [19]. Despite its limited applicability in diverse datasets, the method achieved high sensitivity in average physiological records (98.56% on MITDB, 99.23% on QT) and pathological signals (96.40% on MITDB, 93.07% on BUT PDB).

In another study, an adaptive P wave search approach was developed to identify P waves without prior information automatically. The proposed method was validated only on the MIT-BIH Arrhythmia database (MITDB) and the QT database (QTDB). Only for a few beats does the detection algorithm achieve a sensitivity of 99.96%, a positive predictive value of 99.9%, and an error rate of 0.13% across all validation databases [20]. Similarly, an XGBoost, tree-based, gradient boosting classifier produced the F1 score of 0.8245 for classifying only atrial fibrillation using PhysioNet 2017 Challenge data [21].

Despite attaining good performance as ECG classifiers, these methods have drawbacks, as they may need to capture the dynamic nature and uncertainties involved in the diagnostic process effectively, and they often consider only a limited set of prediction labels. The classification of ECG signals poses challenges due to various issues in the classification process. These include the lack of standardized ECG features, variability among ECG features, the individuality of ECG patterns, the absence of optimal classification rules, and the variability in ECG waveforms among patients [22].

Researchers are exploring alternative approaches that frame diagnostic inferencing as a sequential decision-making process to address these limitations. By incorporating reinforcement learning (RL) [23], they aim to leverage less labeled data alongside relevant evidence from external resources, enabling more informed and accurate diagnoses. RL is a method that depends on goal-directed learning instead of supervised and unsupervised learning. Interacting with the environment and noticing status changes are two ways that learning occurs. Over the past ten years, personalized medicine has significantly benefited from using RL [24].

In recent times, there have been notable advancements in the theoretical and technical aspects of generalization, representation, and efficiency. These developments have paved the way for successfully applying reinforcement learning (RL) techniques, including automated medical diagnosis, in diverse healthcare domains [25].

Recently, reinforcement learning (RL) has been used to detect R waves in electrocardiogram (ECG) signals. The proposed approach involved removing lower frequency components from the signal and identifying the peak candidates using Q learning. The experimental evaluation of the method was conducted using the MIT-BIH dataset, yielding an accuracy of 86.8%. This accuracy is obtained by optimizing the alpha and gamma parameters, which are found to be 0.1 and 0.9, respectively [26]. Another aspect of automated medical diagnosis is a model-driven approach proposed for analyzing electrocardiogram (ECG) signals. A systematic framework was introduced to decompose ECG signals into overlapping lognormal components, and reinforcement learning was utilized to train a deep neural network for estimating the modeling parameters using ECG recordings of infants aged 1 to 24 months [27]. Also, a reinforcement learning-based model was used to automatically fine-tune hyperparameters and network configuration in a CNN model for Arrhythmia prediction from ECG data. The proposed model achieves higher accuracy (97.4%), faster execution time (0.33 mins), and lower mean square error compared to baseline and manually fine-tuned models [28].

### III. CONTRIBUTION OF THIS RESEARCH

This research aims to address the following areas of interest:
1. Exploratory data analysis is conducted to apply exclusion criteria based on age and disease labels. To achieve a more informed diagnosis, it is necessary to separate fetal data from adult patients and Holter device experimental studies from hospital records. Furthermore, an analysis is performed on 133 SNOMED labels to identify disease classes relevant to the P wave and PR interval.
2. Use ECG datasets from diverse demographics to study P wave and PR interval-related variations at the patient level for five different beat types (including normal physiology and pathological beats).
3. Develop a classifier using reinforcement learning that can generalize over all datasets included in the study.
4. Evaluate the ECG classification algorithm based on accuracy execution time and other metrics for multilabel classification.

### IV. METHODS & MATERIALS

The methodology for this research is designed to explore the potential of reinforcement learning on diverse ECG datasets and develop a generalized model that can diagnose P waves and PR interval-related variations in clinical ECG data. The primary research pipeline is given in Fig.1.

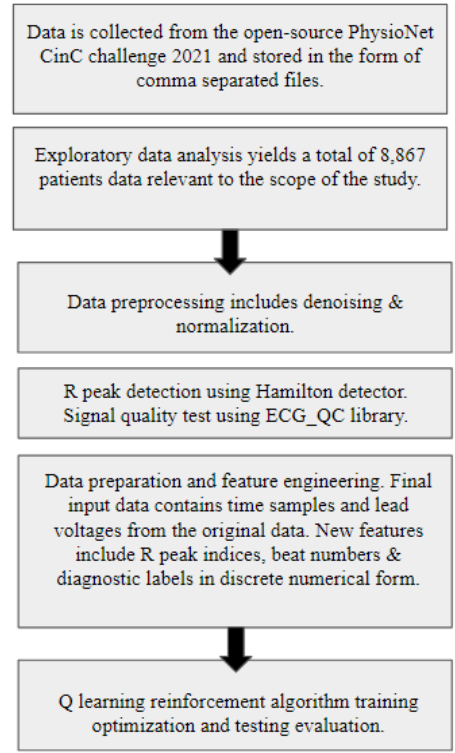

Fig. 1. Research methodology pipeline. Basic steps involve data collection, analysis, and preprocessing. Final input data is classified using Q learning.

#### A. DATA COLLECTION

ECG data is collected from George B. Moody PhysioNet Challenge 2021 [29]. The Challenge data provides annotated twelve-lead ECG recordings from six sources in four countries across three continents. These databases include over 100,000 twelve-lead ECG recordings, with over 88,000 ECGs shared publicly as training data, 6,630 ECGs retained privately as validation data, and 36,266 as test data. This research includes recordings from the six databases given in Table I.

TABLE I
PHYSIONET ELECTROCARDIOGRAM DATASETS

| Dataset | Features |
|---|---|
| CPSC Database and CPSC-Extra Database | 10,330 ECGs, 6 and 144 seconds, sampling frequency of 500 Hz |
| Chapman-Shaoxing and Ningbo Database | 45,152 ECGs, 10 seconds, sampling frequency of 500 Hz |
| The Georgia 12-lead ECG Challenge (G12EC) Database | 10,344 ECGs, 5 and 10 seconds, sampling frequency of 500 Hz |
| PTB and PTB-XL Database | 22,353 ECGs, 10 and 120 seconds, sampling frequency of 500 or 1,000 Hz |

## B. DATA ANALYSIS

The analysis involves six databases and focuses on age, gender, and diagnostic labels recorded using SNOMED [30]. Of the 133 disease labels, 18 are specifically related to the P wave and PR interval region. The dataset consists of 47,433 patients with these 18 labels. From this dataset, five labels were selected based on mono-labeled data, where each patient was assigned a single diagnosis label related to either the P wave or PR interval. Additionally, 679 patient IDs under 18 were excluded from the study, resulting in a final patient count of approximately 8,867. The summary is given in Table II. Table III and IV discuss Waveform variations related to P waves and PR intervals for five relevant classes.

TABLE II
FIVE DISEASE CLASSES AND PATIENT COUNT INCLUDED IN THE RESEARCH AFTER PHYSIONET DATA ANALYSIS

| Diagnostic Label | Total Patients | Mono Labeled Patients | Beat Samples |
|---|---|---|---|
| Normal Sinus Rhythm (NSR) | 28971 | 5355 | 27019 |
| Atrial Fibrillation (AF) | 8374 | 1200 | 27889 |
| Atrial Flutter (AFL) | 5255 | 1465 | 846 |
| Left Atrial Enlargement (LAE) | 1299 | 115 | 1653 |
| 1st Degree AV Block (1AVB) | 3534 | 732 | 14265 |

TABLE III
P WAVE MORPHOLOGY & PR INTERVAL VARIATION IN LEAD II

| Diagnosis | Lead II | |
|---|---|---|
| | P wave | PR Interval |
| NSR | The P wave is less than 120 milliseconds wide and less than 2.5 millimeters high. | 0.12-0.2 seconds |
| AF | Undiscernible P waves | Absent |
| AFL | Saw-toothed false P waves | Absent |
| LAE | Duration is longer than 120 milliseconds. | |
| 1AVB | Normal | PR Interval > 0.2 s |

TABLE IV
P WAVE MORPHOLOGY & PR INTERVAL VARIATION IN LEAD V1

| Diagnosis | Lead V1 | |
|---|---|---|
| | P wave | PR Interval |
| NSR | The P wave is typically biphasic with similar positive and negative deflection sizes. | Normal |
| LAE | > 40ms wide & > 1mm deep. | Normal |

## C. DATA PREPROCESSING

To obtain the most accurate projection of P waves without redundancy, only Lead II and Lead V1 are considered instead of using all twelve leads of the electrocardiogram (ECG). A passband filter is applied with cut-off frequencies ranging from 0.1Hz to 100Hz [19] to eliminate unwanted frequencies, and a digital notch filter is employed to remove 50-60Hz powerline noise [20]. The data is normalized using the Standard Scaler library in Python to ensure consistent scaling. Fig. 2. and Fig. 3 demonstrate the denoising effect on baseline.

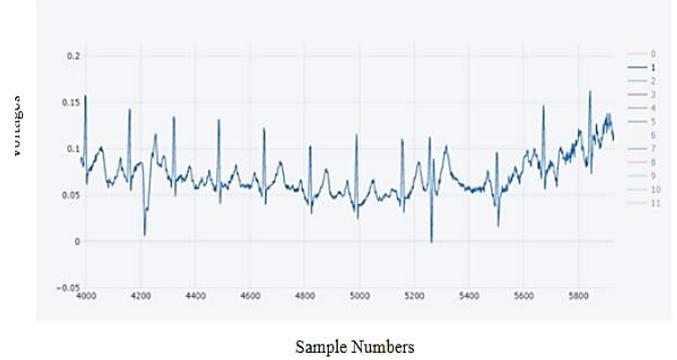

Fig. 2. Plotly graph showing Lead II before denoising. Raw ECG signal has a significant ratio of noises like baseline wander, powerline noise, and artifacts.

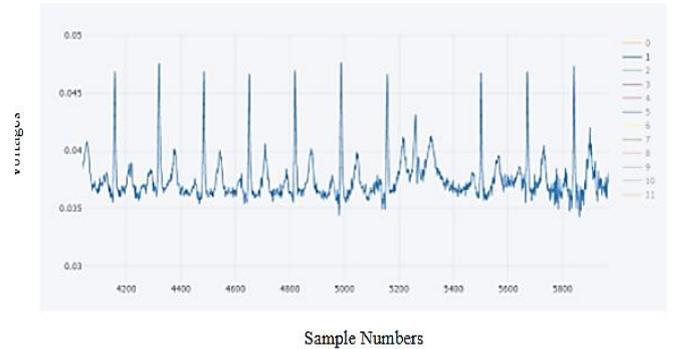

Fig. 3. Plotly graph showing Lead II after denoising. Denoised ECG signal has a significant reduction in signal-to-noise ratio.

## D. LEAD QUALITY

To check the performance of denoising filters on the quality of signals, the ECG_QC library has been used [21]. Bad quality or zero corresponds to an ECG signal containing a baseline shift and high-frequency noise, which disturbs the QRS analysis. Good quality or one corresponds to a clean ECG signal where the QRS can be ideally detected. Table V evaluates the signal quality difference before and after applying the denoising filter. After removing noise, R-R interval variability, skewness, and powerline noise were reduced. Kurtosis has slightly decreased, indicating the peaked distribution of the signal. The baseline power score has also improved on a scale of 0-1, reducing baseline wander noise.

TABLE V
ECG SIGNAL QUALITY SCORE

| Signal Quality Index | Before Denoising | After Denoising |
|---|---|---|
| Variability in the R-R Interval | 0.598 | 0.284 |
| Skewness | 0.104 | 0.032 |
| Kurtosis | -0.316 | -0.583 |
| Power Spectrum | 0.562 | 0.371 |
| Baseline | 0.697 | 0.816 |

### E. FIDUCIAL POINT DETECTION

QRS complexes for Lead II and Lead V1 are detected using Hamilton Segmentor [31]. Output is the sample number (or index value) representing either Q, R, or S peak. Detected peaks are illustrated in Fig. 4.

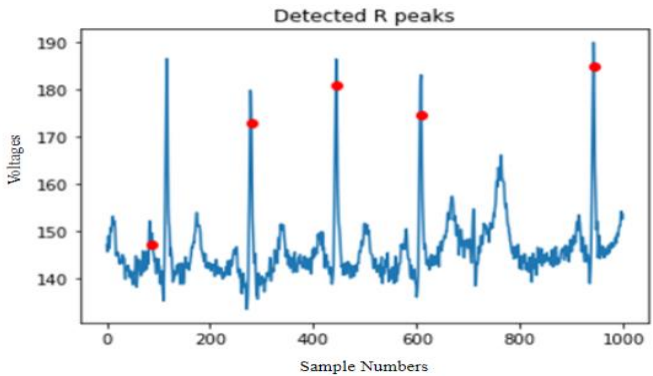

Fig. 4 QRS Complex Peak Detection Using Hamilton Segmentor. Red dots indicate the sample numbers for various R peak indices. *Besides R peaks, the Hamilton QRS detector detects peaks like Q and S.*

### F. REINFORCEMENT LEARNING

Q learning [32] is a model-free off-policy reinforcement algorithm. Implementation is as follows:
1. *Data Representation:* Input data for the algorithm includes CSV columns containing patient IDs, sample numbers, lead voltage data, QRS peak indices, beat number, and diagnostic label integer.
2. *Environment, States, & Actions:* The Q learning environment is an interactive space for the Q learning agent. In this research, the grid world environment is inspired by the electrocardiograph grid paper, as shown in Fig. 5. It represents states and actions related to the ECG data. Dimensions of the grid are decided on the voltage levels between -1 and +1 voltage range and sample numbers. Hence, one frame of the environment is 21xR, where 21 is the voltage level on the Y-axis and R is the peak index representing sample numbers included in one beat. Each beat is a state, and the corresponding diagnostic label is the action. Hence, all beats represent the entire state space of variable sequences, and all five labels represent discrete action space.
3. *Reward, Policy, & Value:* The reward function is designed so the agent receives +1 for accurate classification and -1 for misclassification. Time of classification is deducted from the final reward as a penalty. The confidence score or the maximum Q value is calculated using the epsilon greedy policy, and confidence probability is computed using the SoftMax policy. This score is also added to the final reward. The Q value function shows how good a specific action is for a given state. These values are stored in QSA tables initialized at zero. The five columns of QSA tables represent five discrete actions (labels), and the number of rows represents the number of states (beats).

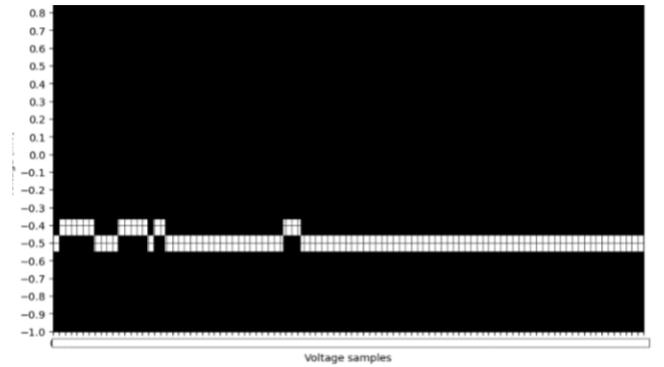

Fig. 5 ECG data representation in a grid world environment. The y-axis represents 21 voltage levels between -1 and +1. The x-axis has sample numbers for one beat. White boxes in the grid represent state coordinates, while black boxes have zero values in the mathematical grid.

4. *Training, Testing, & Periodic Evaluation:* Q agent is trained and optimized by manually tuning hyperparameters like learning rate (alpha), discount factor (gamma), and exploration vs exploitation rate. After optimizing training over 100 episodes, the Q agent is tested for 50 episodes and evaluated based on accuracy, hamming loss, and mean reward increase for every 10th episode.

## V. RESULTS

1. *Training Convergence:* The Q Learning algorithm was trained with an increment of 10 episodes or iterations until convergence was achieved at the 100th episode. Figures 6 and 7 demonstrate gradual reward accumulation from the first episode to episodes 5 and 10. When convergence is reached, the Q agent does not explore new information from the data, and its performance becomes stable.

2. *Comparison of Reward Functions:* The Q agent was tested on three reward functions. A simple reward function included +1 for accuracy and -1 for classification inaccuracy. The second-degree reward had a time penalty where time for execution was deducted from the final reward. The degree reward function contained Qmax values regarding epsilon greedy policy or confidence score probability in SoftMax. These values are added as a positive reward for the Q agent. The performance of agents on different reward definitions is shown in Table VI. Regarding accuracy and time penalty, the best-performing reward function is the third-degree reward function on SoftMax policy, with an average confidence score of 0.39. The confidence score value is moderate because our Q learning algorithm

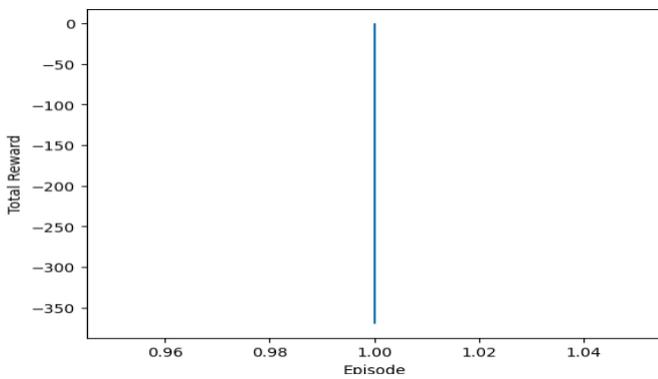

produces Q tables for individual patients where each beat is mapped to a single label.

Fig. 6 No reward in episode 1.00. A vertical blue indicates zero reward in the first episode, saying no actions have been taken by the Q agent yet.

3. *Hyperparameter Optimization:* The best performance of the Q agent is obtained at 0.001 learning rate and 0.9 discount factor. Average accuracies obtained by changing hyperparameters are shown in Fig.8 and 9. A 0.001 alpha value enables slow learning, and the agent can explore the environment exhaustively. The discount factor of 0.9 promotes agents' behavior to focus on future rewards and be vigilant of the consequences of their actions. Tables VII & VIII compare Q agent performance on two discount factor values at a constant alpha rate 0.001. At 0.1 gamma value, the confidence score decreases, showing the agent is least sure about its actions as it is focused on accumulating immediate rewards.
4. Testing & Evaluation: Training and testing rewards gradually increase over episodes and become constant after specific iterations, as shown in Fig. 9. Average reward increases during the testing phase, indicating that the Q agent is performing well. Hamming loss calculates the fraction of incorrectly predicted or missing labels compared to the actual labels. As shown in Fig. 10. Average accuracies increase while hamming loss decreases over every ten-episode interval. Table XI evaluates Q agents tested after training on 100 episodes. Average accuracy is 90.4%, while hamming loss is 9.6%. The evaluation metrics [33] used for the multilabel classification by Q-Agent are given below.

    a. *Accuracy:*
(Number of correctly classified beats) / (Total number of beats)
    b. *Precision:*
(Number of True positives) / (Number of True positives + Number of false positives)
    c. *Recall:*
(Number of true positives for the class) / (Number of true positives + Number of false negatives for the class)
    d. *F1 Measure:*
2 * (Precision * Recall) / (Precision + Recall)
    e. *Hamming Loss:*
(Number of misclassified labels) / (Total number of labels)

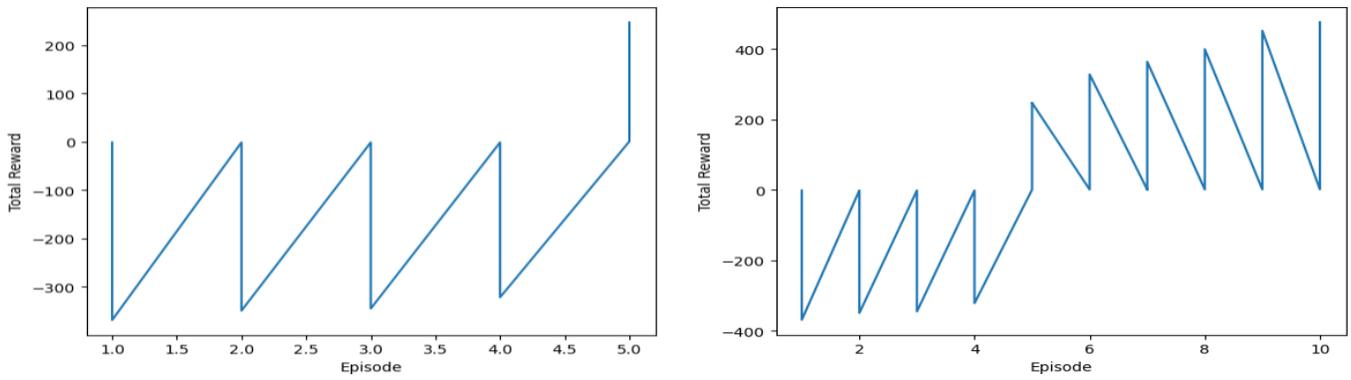

Fig. 7 Reward accumulation gradually increases to positive values from episode 5.00 (left) to episode 10.00 (right)

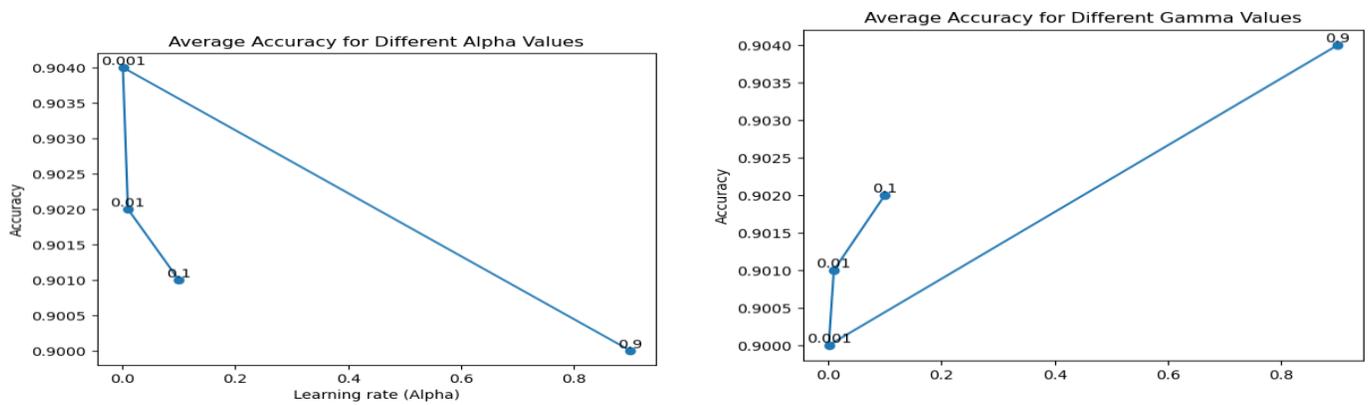

Fig. 8. Hyperparameters tuning for optimizing Q agent. The average accuracy for all five classes is calculated for different values of alpha and gamma. The highest average accuracy is 90.4%, with 0.01 alpha and 0.9 gamma values.

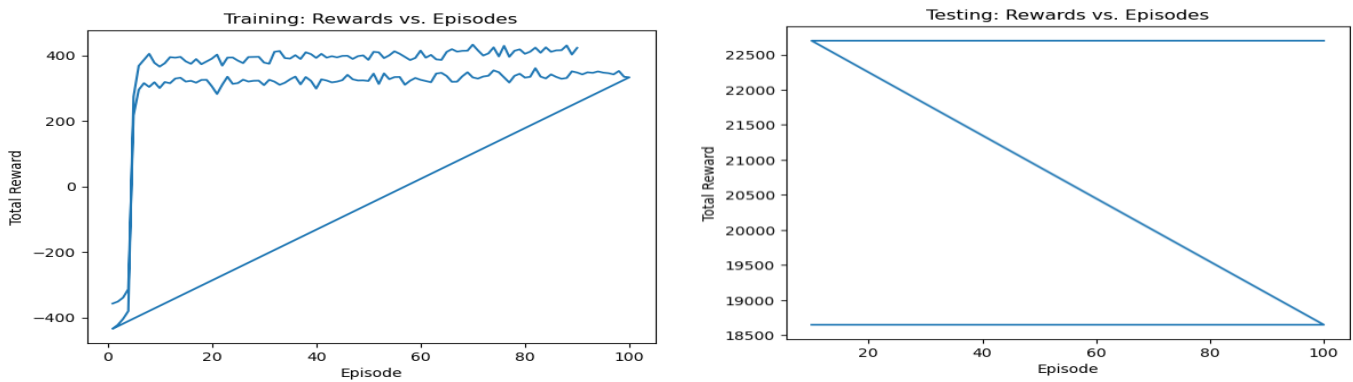

Fig. 9. Training and Testing rewards. In training (left), the agent gradually accumulates rewards by exploring the environment. During testing reward (right), the agent exploits the information it has learned, giving a smooth graph.

TABLE VI
PERFORMANCE OF Q LEARNING AGENT ON VARIOUS REWARD FUNCTION

| **Reward in the 100th Episode using Epsilon Greedy Policy** | | | | | |
|---|---|---|---|---|---|
| Reward Definition | ECG Classes | Total Reward | Accuracy | Time of Classification (seconds) | Confidence Value ($Q_{max}$) |
| Reward based on *accuracy* | NSR (0) | 396 | 0.918 | | |
| | AF (1) | 435 | 0.9112 | | |
| | AFL (2) | 98 | 0.901 | | |
| | LAE (3) | 254 | 0.891 | | |
| | 1AVB (4) | 326 | 0.884 | | |
| Average | | 361.8 | 0.9012 | | |
| Reward based on *accuracy & time penalty* | NSR (0) | 209.3654625 | 0.924 | 0.0606 | |
| | AF (1) | 404.1954193 | 0.9122 | 0.1797 | |
| | AFL (2) | 270.0528943 | 0.902 | 0.1371 | |
| | LAE (3) | 225.8052170 | 0.889 | 0.0332 | |
| | 1AVB (4) | 296.0369725 | 0.884 | 0.0943 | |
| Average | | 281.4911931 | 0.90212 | 0.10198 | |
| Reward based on *confidence value or Qmax, accuracy & time penalty* | NSR (0) | 2649.622210 | 0.92 | 0.1181 | 3.24 |
| | AF (1) | 3163.511501 | 0.91 | 0.1185 | 3.16 |
| | AFL (2) | 766.3391647 | 0.90 | 0.0550 | 3.29 |
| | LAE (3) | 1327.599108 | 0.89 | 0.0644 | 3.18 |
| | 1AVB (4) | 306.4352185 | 0.88 | 0.0186 | 3.15 |
| Average | | 1642.301040 | 0.900 | 0.07412 | 3.204 |
| **Reward in 100th Episode using SoftMax Policy** | | | | | |
| Reward Definition | ECG Classes | Total Reward | Accuracy | Time of Classification (seconds) | Confidence Probability Score |
| Reward based on *confidence probability score, accuracy & time penalty* | NSR (0) | 348.7888924 | 0.92 | 0.0683 | 0.39 |
| | AF (1) | 465.6657739 | 0.91 | 0.1097 | 0.39 |
| | AFL (2) | 112.0626190 | 0.91 | 0.0160 | 0.39 |
| | LAE (3) | 182.6409696 | 0.89 | 0.0258 | 0.39 |
| | 1AVB (4) | 611.1028246 | 0.89 | 0.015 | 0.39 |
| Average | | 274.31449928 | 0.904 | 0.06236 | 0.39 |

TABLE VII
BEST PERFORMANCE OF Q LEARNING AGENT BY OPTIMIZING ALPHA & GAMMA VALUES

| Learning rate | Discount factor | ECG Class | Total Reward | Accuracy | Time of Classification (seconds) | Confidence Probability Score |
|---|---|---|---|---|---|---|
| 0.001 | 0.9 | NSR (0) | 348.7888924 | 0.92 | 0.0683 | 0.39 |
| | | AF (1) | 465.6657739 | 0.91 | 0.1097 | 0.39 |
| | | AFL (2) | 112.0626190 | 0.91 | 0.0160 | 0.39 |
| | | LAE (3) | 182.6409696 | 0.89 | 0.0258 | 0.39 |
| | | 1AVB (4) | 265.4142415 | 0.89 | 0.0920 | 0.39 |
| Average | | | 274.3144992 | 0.904 | 0.06236 | 0.39 |

TABLE VIII
BEST PERFORMANCE OF Q LEARNING AGENT BY OPTIMIZING ALPHA & GAMMA VALUES

| Learning rate | Discount factor | ECG Class | Total Reward | Accuracy | Time of Classification (seconds) | Confidence Probability Score |
|---|---|---|---|---|---|---|
| 0.001 | 0.1 | NSR (0) | 365.1344810 | 0.92 | 0.0927 | 0.04 |
|  |  | AF (1) | 480.7137572 | 0.91 | 0.0710 | 0.04 |
|  |  | AFL (2) | 109.0254470 | 0.91 | 0.0262 | 0.04 |
|  |  | LAE (3) | 259.1329067 | 0.89 | 0.0542 | 0.04 |
|  |  | 1AVB (4) | 278.3072589 | 0.88 | 0.0841 | 0.04 |
| Average |  |  | 298.46277016 | 0.902 | 0.06564 | 0.04 |

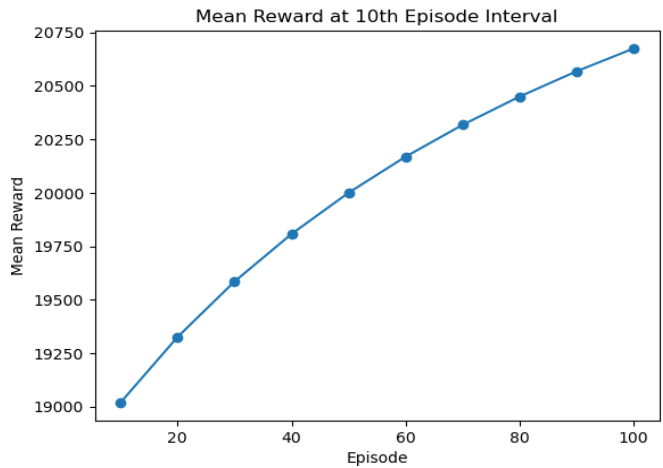
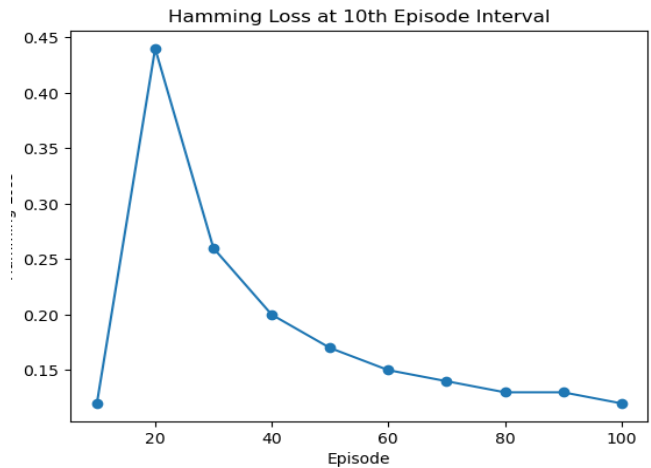

Fig.10 Mean reward increase (left) during periodic evaluation shows that the Q agent performs well in the testing phase. Hamming loss (right) decreases, leading the Q agent to make fewer errors in predicting ECG labels during the testing phase.

TABLE IX
AGENT TESTING AFTER 100$^{TH}$ EPISODE

| Evaluation Metrics During Q Agent Testing after the 100$^{th}$ Episode | | | | | |
|---|---|---|---|---|---|
| ECG Class | Accuracy | Precision | Recall | F1 | Hamming Loss |
| NSR (0) | 0.92 | 0.92 | 0.92 | 0.92 | 0.12 |
| AF (1) | 0.91 | 0.91 | 0.91 | 0.91 | 0.12 |
| AFL (2) | 0.91 | 0.91 | 0.91 | 0.91 | 0.12 |
| LAE (3) | 0.89 | 0.89 | 0.89 | 0.89 | 0.12 |
| 1AVB (4) | 0.89 | 0.89 | 0.89 | 0.89 | 0.12 |
| Average | 0.904 | 0.904 | 0.904 | 0.904 | 0.12 |

## VI. Conclusion

This research demonstrates the effectiveness of reinforcement learning, specifically Q learning, in ECG classification without explicit instructions or predefined rules. The problem of ECG classification was formulated as an off-policy reinforcement learning task, leading to promising results in accurately classifying diverse datasets from different regions. The approach proved to be efficient in dealing with imbalanced data. Interestingly, the Q agent classified atrial flutter and atrial fibrillation with 91% accuracy, though these classes are similar and pattern differentiation is difficult. Q agent reinforcement learning has a faster average execution or classification time of 0.04 seconds compared to profound learning studies where execution time exceeds 10 seconds (0.33 minutes, as shown in the background section). This is due to the efficient reward system with a time penalty designed for Q agents to make quick diagnoses. Unlike traditional machine learning methods, the Q learning agent considers diversity at demographic and patient levels, effectively handling variations related to P wave and PR interval with higher accuracy and robustness. This approach allows for a more comprehensive and reliable classification of ECG data compared to other existing methods.

## Acknowledgment

The authors would like to express their sincere gratitude to Prof. Dr. Shahzad Younis and Dr. Mehak Rafiq, whose invaluable insights and guidance significantly enriched this research. We also thank Dr. Shahzad Rasool for his continuous support throughout this research.